\documentclass[conference]{IEEEtran}
\IEEEoverridecommandlockouts
\usepackage{cite}
\usepackage{amsmath,amssymb,amsfonts}
\usepackage{graphicx}
\usepackage{textcomp}
\usepackage{xcolor}
\usepackage{mathtools}
\usepackage{verbatim}
\usepackage{enumitem}
\usepackage{booktabs} 
\usepackage[skip=1pt]{caption}
\usepackage{multirow}
\usepackage[a4paper, total={184mm,239mm}]{geometry}
\usepackage{fancyhdr}

\def\BibTeX{{\rm B\kern-.05em{\sc i\kern-.025em b}\kern-.08em
    T\kern-.1667em\lower.7ex\hbox{E}\kern-.125emX}}
\begin{document}
\title{MEMHD: Memory-Efficient Multi-Centroid Hyperdimensional Computing for Fully-Utilized In-Memory Computing Architectures}

\author{
    \IEEEauthorblockN{Do Yeong Kang\textsuperscript{1}, Yeong Hwan Oh\textsuperscript{1}, Chanwook Hwang\textsuperscript{1}, Jinhee Kim\textsuperscript{1}, Kang Eun Jeon\textsuperscript{1}$^{*}$\thanks{$^{*}$: Corresponding authors} and Jong Hwan Ko\textsuperscript{1, 2}$^{*}$}
    \IEEEauthorblockA{\textsuperscript{1}Department of Electrical and Computer Engineering, Sungkyunkwan University, Suwon, South Korea\\
    \textsuperscript{2}College of Information and Communication Engineering, Sungkyunkwan University, Suwon, South Korea}
    \IEEEauthorblockA{Email: \{ksps7106, yh991111, ghkdcks12, a2jinhee, kejeon, jhko\}@skku.edu}
}

\maketitle

\begin{abstract}
The implementation of Hyperdimensional Computing (HDC) on In-Memory Computing (IMC) architectures faces significant challenges due to the mismatch between high-dimensional vectors and IMC array sizes, leading to inefficient memory utilization and increased computation cycles. This paper presents MEMHD, a Memory-Efficient Multi-centroid HDC framework designed to address these challenges. MEMHD introduces a clustering-based initialization method and quantization-aware iterative learning for multi-centroid associative memory. Through these approaches and its overall architecture, MEMHD achieves a significant reduction in memory requirements while maintaining or improving classification accuracy. Our approach achieves full utilization of IMC arrays and enables one-shot (or few-shot) associative search. Experimental results demonstrate that MEMHD outperforms state-of-the-art binary HDC models, achieving up to 13.69\% higher accuracy with the same memory usage, or 13.25x more memory efficiency at the same accuracy level. Moreover, MEMHD reduces computation cycles by up to 80x and array usage by up to 71x compared to baseline IMC mapping methods when mapped to 128x128 IMC arrays, while significantly improving energy and computation cycle efficiency.
\end{abstract}


\section{Introduction}
HDC is an emerging computational paradigm inspired by the brain's distributed processing and the properties of high-dimensional vector spaces \cite{kanerva}. By representing data as high-dimensional vectors, known as hypervectors, HDC enables efficient and robust computation suitable for various machine learning tasks. The high dimensionality provides inherent robustness to noise and facilitates operations such as binding and superposition, which are analogous to logical operations in traditional computing. Applications of HDC span a wide range, including language processing\cite{language}, robotics\cite{robotics}, and biosignal classification\cite{biosignal}.

\begin{figure}[t]
\centering
\includegraphics[width=0.9\linewidth]{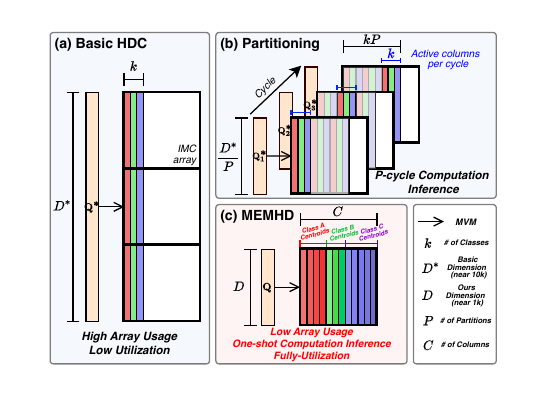}
\caption{Overview of MEMHD}
\label{fig:1}
\vspace{-20pt}  
\end{figure}

IMC has emerged as a promising approach to overcome the von Neumann bottleneck by performing computations directly within memory arrays \cite{imc1}. By integrating processing capabilities into memory units, IMC architectures dramatically reduce data movement, leading to enhanced computational speed and energy efficiency for data-intensive applications. Recent advancements have demonstrated the potential of IMC in accelerating machine learning algorithms and neural networks, utilizing various memory technologies such as SRAM, DRAM, and non-volatile memories like RRAM and PCM \cite{imc2,isaac}.

The synergistic integration of HDC and IMC has become a focal point of recent research, aiming to leverage their complementary strengths. Implementing HDC algorithms on IMC architectures accelerates high-dimensional computations while exploiting the parallelism and energy efficiency inherent in IMC. Various IMC-based approaches for HDC have been explored, including analog and digital processing-in-memory structures \cite{exploring}, PCM-based crossbar arrays \cite{imhdc}, and scalable subarray architectures using emerging devices such as FeFETs \cite{fefet}. These diverse implementations demonstrate the potential of IMC to significantly enhance HDC performance and energy efficiency across different hardware platforms.

Nevertheless, mapping vectors onto IMC arrays presents significant challenges due to dimensional mismatches, because HDC utilizes high-dimensional vectors to achieve noise resilience and effective pattern separation. This often results in underutilization of computational resources, as vector dimensions frequently exceed row dimensions of individual IMC arrays in Fig. \ref{fig:1}-(a). To address this, \textit{partitioning} has been proposed in \cite{imhdc}, dividing hypervectors into smaller segments mapped across previously unused columns in fewer arrays. While this helps fit computations within hardware constraints, it doesn't reduce the total number of computation cycles required for inference, as shown in Fig. \ref{fig:1}-(b). The root cause of these issues lies in the fundamental mismatch between HDC's requirements and IMC array structures. On the one hand, HDC relies on high-dimensional vectors that often exceed the row dimensions of IMC arrays. On the other hand, the traditional HDC approach of using a single vector per class results in most columns of the IMC array remaining unused during similarity calculations, causing column underutilization. To address these challenges, we need an approach tailored to the IMC array structure. Such a method would utilize vector dimensions matching the IMC array's row dimension, while employing a number of class vectors corresponding to its column dimension. 

Building on this insight, we move beyond the HDC's standard associative memory structure where each class is represented by a single class vector. We propose MEMHD, a novel Memory-Efficient Multi-centroid HDC framework that achieves fully-utilized associative memory while significantly reducing the dimensionality of hypervectors and maintaining classification accuracy. To the best of our knowledge, this is the first attempt to develop a multi-centroid model for HDC's associative memory specifically designed to fit IMC array sizes. To effectively train the multi-centroid associative memory, we present two key phases. First, we employ clustering-based initialization methods for initial centroid placement, enabling more stable training that leads to higher accuracy and faster convergence. Second, we utilize iterative learning techniques that account for quantization effects during the training process. MEMHD substantially reduces memory requirements for both encoding and associative search while achieving comparable or superior accuracy to existing binary HDC baselines. This is accomplished by using dimensions compatible with IMC arrays (near 1k), which are significantly smaller than the 10k dimensions commonly used in existing HDC approaches. Therefore, our model achieves 13.69\% higher accuracy compared to the baselines with the same memory requirements, and $13.25\times$ more efficient memory usage for the same level of accuracy. Compared to the IMC baselines, shown in Fig. \ref{fig:1}, MEMHD reduces computation cycles by $80\times$ and array usage by $71\times$.

\section{Background : HDC Classification}
\subsection{Overall HDC Structure}
The HDC framework is structured around two primary modules: the encoding module (EM) and the associative memory (AM). During training, the EM processes the input data to extract relevant features and converts them into hypervectors. The encoded hypervectors are used to create class vectors, each representing a specific class, which are then stored in the AM for future reference. In runtime, new data is encoded into a hypervector using the same method, and an associative search is performed to compare this query hypervector against the class vectors in the AM using similarity measures. 

\subsection{Hypervector Encoding}
Random Projection encoding involves Matrix-Vector-Multiplication (MVM) between the $f\times D$ projection matrix $M$ and the $f$-dimensional input feature vector $F$. Each column of $M$ consists of a randomly generated base vector $B$, which can be represented as either binary\cite{locality} or floating-point data type\cite{theoretical}. The encoding is performed as follows:
\begin{equation}
    H = M^\top F.
\end{equation}
Another approach is ID-Level encoding, where each input feature is associated with a random binary ID hypervector, and each feature value is represented by a Level hypervector. The input is encoded by element-wise multiplication of ID and Level hypervectors at each position, then summing across all positions: $H = \sum_{i=1}^{f} (ID_i \otimes L_{x_i})$, where $ID_i$ is the ID hypervector for the $i$-th position, and $L_{x_i}$ is the Level hypervector for the value $x_i$.

\subsection{Associative Memory Training}
In HDC, it is possible to learn class vectors in a single pass. This is achieved by constructing the class vector through the summation of all sample hypervectors associated with the same class, as follows: $C_k = \sum_{\forall i} H^{i}_{k}$, where $C_k$ is the class vector for class $k$, and $H^{i}_{k}$ are the sample hypervectors with label $k$.

Iterative learning updates the AM by focusing exclusively on the hypervectors that were misclassified across the entire training dataset. For each mispredicted hypervector $H$, the true class vector is adjusted to be more similar to $H$, while the predicted class vector is adjusted to be less similar. This approach enables the model to improve its predictions over several iterations, and can be summarized as follows:
\begin{equation}
C_{true} = C_{true} + \alpha H ,\quad C_{pred} = C_{pred} - \alpha H
\end{equation}
where $\alpha$ is the learning rate.
For the binary HDC model, the resulting $C$ is binarized to produce a binary class hypervector.

\subsection{Associative Search}
For classification tasks, associative search involves selecting a predicted class based on the similarity between a query vector and class vectors. The class with the highest similarity to the query vector is inferred as the predicted class. A primary similarity metric used in this context is dot similarity:
\begin{equation}
\delta_{\text{dot}}(A, B) = A \cdot B
\end{equation}
where A and B are vectors of the same dimension.
While various similarity measures such as Hamming distance and cosine similarity exist for associative search, dot similarity is a simple yet effective measure of vector similarity. It aligns well with the multiplication and accumulation operations inherent in IMC arrays.

\begin{figure*}[t]
\centering
\includegraphics[width=0.95\textwidth]{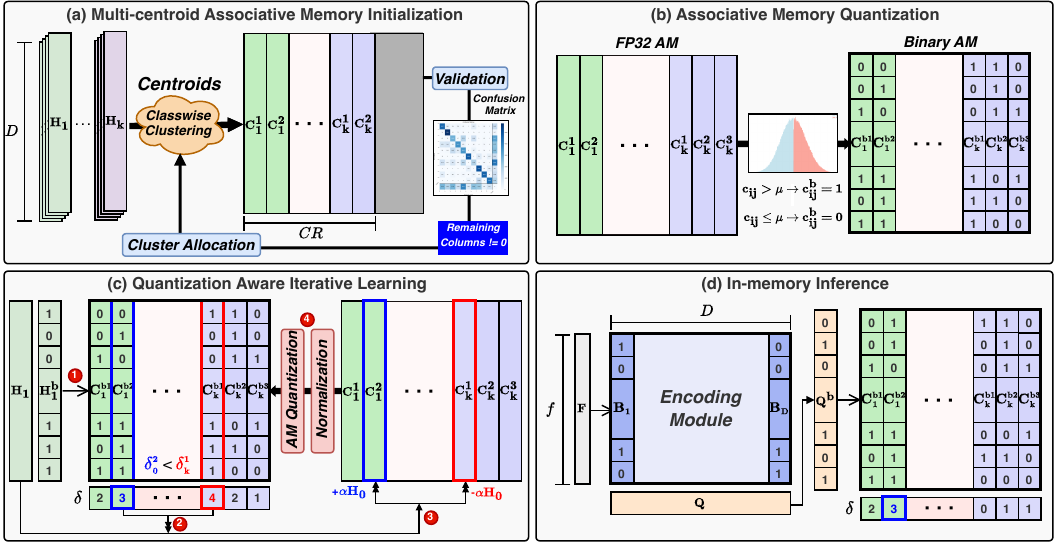}
\caption{Overall MEMHD framework}
\label{fig:2}
\vspace{-18pt}
\end{figure*}

\section{MEMHD: Memory-Efficient Multi-Centroid HD}
In this section, we introduce the overall method for our multi-centroid AM. To address the time-consuming associative search in existing partitioning methods, we employ an IMC array-optimized multi-centroid AM structure. This approach enables one-shot (or few-shot) associative search and utilizes hypervectors with dimensions tailored to the IMC array size, significantly reducing encoding costs. Fig. \ref{fig:2} illustrates the overall framework of MEMHD. Our multi-centroid AM is constructed using the clustering-based initialization (shown in Fig. \ref{fig:2}-(a), referred to in \ref{A}), which enhances accuracy and convergence. For the binary AM, we perform 1-bit quantization of the AM (shown in Fig. \ref{fig:2}-(b), referred to in \ref{B}) and subsequently refine the model through quantization-aware iterative learning (shown in Fig. \ref{fig:2}-(c), referred to in \ref{C}), optimized for efficient IMC mapping. Considering direct mapping to IMC array, we utilize the projection encoding and dot similarity for associative search, which are based on MVM (shown in Fig. \ref{fig:2}-(d), referred to in \ref{D}).

\subsection{Multi-Centroid AM Initialization}
\label{A}
In traditional HDC approaches, each class is represented by a single vector. The initialization method for these class vectors, including random sampling, has minimal impact on their ability to effectively represent the overall set of sample hypervectors for that class. Random sampling is commonly used for initialization due to the method's robustness to initial conditions. During the subsequent iterative learning process, all updates corresponding to a single class are applied to one class vector, allowing it to converge to a representative vector for the class with minimal dependence on its initial state. 

In contrast, multi-centroid models use multiple class vectors per class, each capturing distinct features of that class and learning independently. This makes the initialization method critical for ensuring comprehensive class representation. Randomly-sampled initial class vectors fail to distribute the centroids evenly across point cloud of the class sample hypervectors, leading to poor representation. To address this, we propose the clustering-based initialization method that improves convergence and accuracy.

1) \textit{Classwise Clustering} : First of all, as shown in Fig. \ref{fig:2}-(a), MEMHD splits the encoded sample hypervectors by class and applies clustering for each class. We use K-means algorithm for clustering, which allows us to specify the number of clusters directly. The distance metric used in clustering is based on dot similarity, the same metric employed in associative search. This consistency ensures that the clustering process is optimized for subsequent associative search operations. We define a hyperparameter $R$ as the proportion of columns used for initial clustering out of the total memory columns. This approach allows for cluster allocation, which is detailed in (2). Given $R$, the number of initial clusters $n$ per class is determined by the equation: $n = \text{max}\left(1, \left\lfloor{\frac{\mathit{CR}}{k}}\right\rfloor\right) \label{eq}$, where $C$ is the total number of columns and $k$ is the number of classes. After clustering, each cluster's centroid becomes an initial class vector.

2) \textit{Cluster Allocation} : We allocate the remaining columns, $C$(1-$R$), based on a validation process using the entire training dataset. This process employs a confusion matrix to assess the distribution of misclassifications among different classes. Classes with higher misclassification rates are assigned additional centroids to enhance their representation in the model. These additional centroids are allocated to classes with higher misprediction rates. This process of validation, cluster assignment, and re-clustering is repeated until no remaining columns are left. Once all columns are utilized, resulting in a fully utilized IMC array, the initialization process is complete.

\subsection{AM Quantization}
\label{B}
To optimize memory efficiency for mapping AM into IMC array, we represent the AM in binary format. The initial AM, initialized with floating-point (FP) values after clustering, exhibits a distribution resembling a Gaussian curve, as illustrated in Fig. \ref{fig:2}-(b). MEMHD performs 1-bit quantization using the mean value as a threshold, binarizing the AM by setting values greater than the mean $\mu$ to 1 and the rest to 0.

\subsection{Quantization-Aware Iterative Learning}
\label{C}
The binary AM is trained through quantization-aware iterative learning, which optimizes the model while accounting for quantization effects. While \cite{quanthd} introduced quantization-aware learning for basic HDC, our multi-centroid model extends this approach with crucial modifications. Specifically, our method introduces additional considerations for setting learning targets and update strategies, tailored to the unique requirements of the multi-centroid structure. Our learning process is carried out in four steps, as illustrated in Fig. \ref{fig:2}-(c):

1) \textit{Dot Similarity}: For each training data sample hypervector, the dot product similarity with the binary AM is evaluated. Updates are triggered only when a misprediction occurs.

2) \textit{Update Target Selection}: A misprediction indicates that the predicted class (with the highest similarity score) is incorrect and that its similarity score exceeds similarities with the true class vectors. The mispredicted class vector with the highest similarity score is designated as the target for update: 
\begin{equation}
(l', m) = \arg\max_{i, j} \; \delta_{\text{dot}} ( C^{bi}_{j} , H^{b}_{l} )
\end{equation}
where $l'$ denotes the mispredicted class, $m$ represents the sub-label with the highest similarity, $j$ indicates the class index, and $i$ denotes the sub-label index.
Ideally, the class vectors within the same class should represent different features, ensuring that each sample vector influences a unique class vector. Therefore, for the true class, the target for update is determined as follows:
\begin{equation}
(l, n) = \arg\max_{i} \; \delta_{\text{dot}} ( C^{bi}_{l} , H^{b}_{l} )
\end{equation}
where $l$ is the true class and $n$ indicates the sub-label of the class vector with the highest similarity among those within the same class.

3) \textit{Iterative Learning}: The iterative learning process is carried out on the FP AM based on the selected update targets: 
\begin{equation}
    C^{n}_{l} = C^{n}_{l} + \alpha H_{l} ,\quad C^{m}_{l'} = C^{m}_{l'} - \alpha H_{l}.
\end{equation}
The learning rate typically ranges from 0.01 to 0.1, depending on the dataset and hyperparameters $D$, $C$. The learning rate is adjusted based on dataset complexity and model size: a lower one is used for more challenging datasets, while a higher one is applied to models with large D or C.

4) \textit{Binary AM Update}: Following vector operations on the FP AM, we perform a normalization step. This normalization, distinct from standard HDC approaches, ensures an even distribution of learning influence across multiple class vectors within the same class, preventing any single vector from dominating. The process concludes with an update to the binary AM, achieved by binarizing the normalized FP AM.

\subsection{In-Memory Inference}
\label{D}
For in-memory inference, both the binary projection matrix as EM and the binary AM are mapped into the IMC array. The inference process utilizes the MVM-based encoding and associative search methods mentioned in Eq. (3): $pred = \arg\max_{i, j} \delta_{\text{dot}} ( C^{bi}_{j}, H^{b} )$.

\section{EVALUATION}
\subsection{Experimental Setup}
Table \ref{tab:1} shows the characteristics, encoding, and training methods of baseline binary HDC models. SearcHD \cite{searchd} is the multi-model structure most similar to our approach, utilizing N-vector quantization, where a non-binary class vector is quantized into multiple binary class vectors. In our evaluation, we fixed N = 64. QuantHD \cite{quanthd} was the first to introduce the quantization-aware learning. LeHDC \cite{lehdc}, based on BNN, is known as the state-of-the-art binary HDC model. All three methods use ID-level encoding with L = 256. In addition to these models, we introduce BasicHDC separately because both its encoding and associative search can be performed using MVM operations, making it directly comparable to IMC array implementations. In contrast, the other baselines use ID-Level encoding, which is not directly compatible with IMC array computations. For IMC baselines, we compare MEMHD with two existing structures in Fig. \ref{fig:1}. Our comparison focuses on (1) the required computation cycles for a single IMC array, the number of arrays needed to map the entire structure, and AM utilization, as well as (2) energy consumption and cycles of AM. Those read and write cycles and energy consumption data are derived from SRAM-based IMC arrays simulated using NeuroSim simulator \cite{neurosim}, as presented in \cite{waam}. Our evaluation employs three datasets: MNIST \cite{mnist}, Fashion-MNIST (FMNIST) \cite{fmnist}, and ISOLET \cite{isolet}. MEMHD are trained for 100 epochs following initialization. All experiments were conducted with 5 trials, and the average accuracy was reported.

\begin{table}[t]
\centering
\caption{Memory requirements of baseline HDC models}
\renewcommand{\arraystretch}{1}
\resizebox{\linewidth}{!}
{
\begin{tabular}{|c|c|c|c|} 
\hline
\multicolumn{1}{|c|}{\textbf{Baseline}} & \multicolumn{1}{c|}{\textbf{Keywords}} & \multicolumn{1}{c|}{\textbf{Encoding}} & \multicolumn{1}{c|}{\textbf{Associative}} \\ 
\multicolumn{1}{|c|}{\textbf{Binary HDC}} & \multicolumn{1}{c|}{\textbf{ \& Methods} }  & \multicolumn{1}{c|}{\textbf{Module}}   & \multicolumn{1}{c|}{\textbf{Memory}}      \\ \hline
$\textbf{SearcHD\cite{searchd}}$ & $\begin{array}{c}\text{Multi-model}\\ \text{ID-Level}\\ \text{Single-pass}\end{array}$ & ${(f + L) \times D}$ & ${k \times D \times N}$ \\ \hline
$\textbf{QuantHD\cite{quanthd}}$ & $\begin{array}{c}\text{ID-Level}\\ \text{Quantization Aware}\\ \text{Iterative-pass}\end{array}$ & ${(f + L) \times D}$ & ${k \times D}$ \\ \hline
$\textbf{LeHDC\cite{lehdc}}$ & $\begin{array}{c}\text{ID-Level}\\ \text{BNN based Training}\end{array}$ & ${(f + L) \times D}$ & ${k \times D}$ \\ \hline
$\textbf{BasicHDC}$ & $\begin{array}{c}\text{Projection}\\ \text{Single-pass}\end{array}$ & ${f \times D}$ & ${k \times D}$ \\ \hline
$\textbf{MEMHD}$ & $\begin{array}{c}\text{Multi-centroid}\\ \text{Projection}\\ \text{Quantization Aware}\\ \text{Iterative-pass}\end{array}$ & ${f \times D}$ & ${C \times D}$ \\ \hline
\end{tabular}
}
\caption*{\small{Note: $f$: \# of features, $L$: \# of levels, $D$: dimensionality, $k$: \# of classes, $C$: \# of columns and $N$: vector quantization factor.}}
\label{tab:1}
\vspace{-18pt}
\end{table}

\begin{figure*}[ht]
\centering
\includegraphics[width=0.93\linewidth]{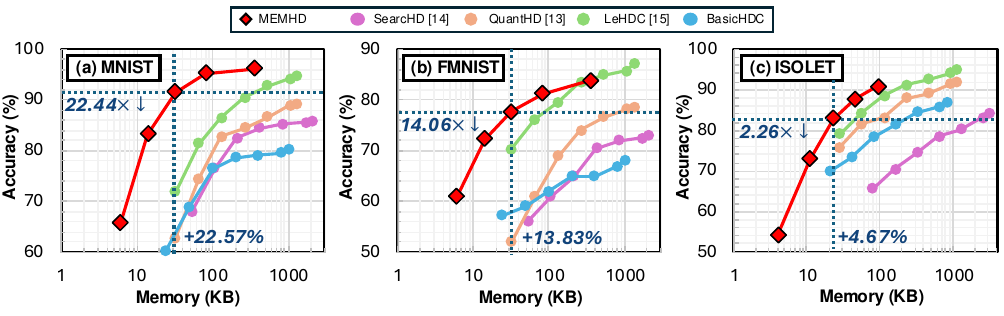}
\caption{Accuracy and memory requirement (KB). MEMHD used for (a) MNIST (b) FMNIST: 64x64 to 1024x1024 square sizes ($D$x$C$), (c) ISOLET: fixed 128 columns, varied dimensions. We set dimensions of baslines from 256D to 10240D.}
\vspace{-14pt}
\label{fig:3}
\end{figure*}

\begin{figure*}[ht]
\centering
\includegraphics[width=0.93\linewidth]{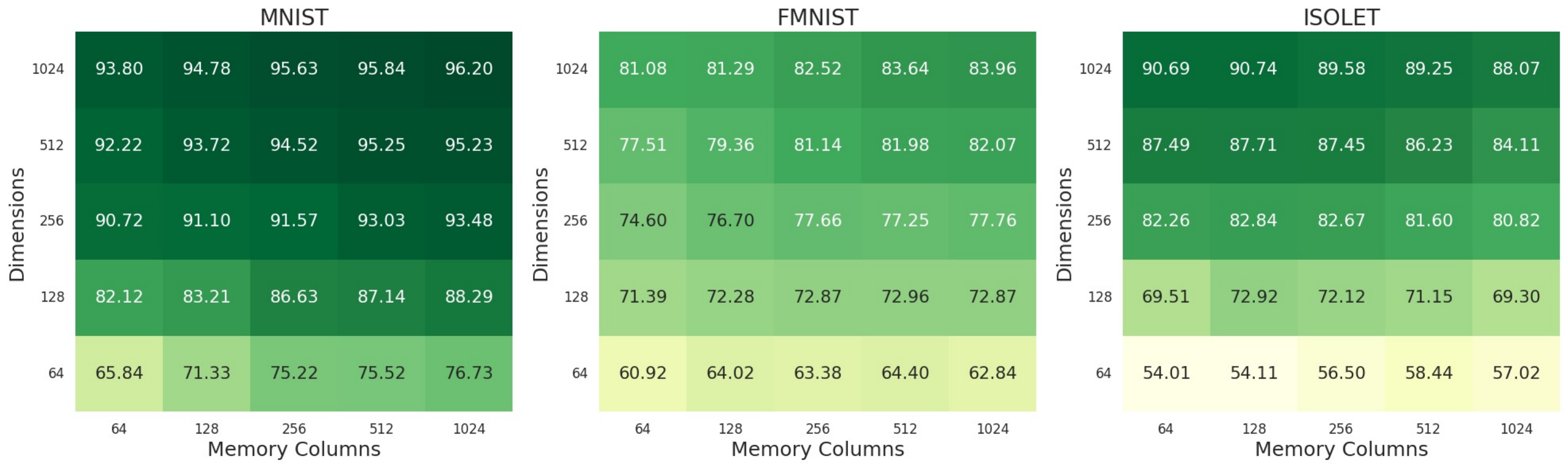}
\caption{MEMHD Accuracy Heatmap from 64x64 to 1024x1024.}
\label{fig}
\vspace{-17pt}
\label{fig:4}
\end{figure*}

\subsection{Accuracy and Memory Requirement}
Fig. \ref{fig:3} compares the accuracy and memory requirement of our model with the baselines for each dataset. Our model demonstrates accuracy close to or exceeding state-of-the-art performance while using significantly less memory. Notably, for MNIST using the 512x512, our approach achieves an accuracy 1.1\% higher than the state-of-the-art while being $16.2\times$ more memory-efficient. Furthermore, our mid-range models (256x256 for MNIST, FMNIST and 256x128 for ISOLET) outperform comparable baselines such as LeHDC and QuantHD. On average, these models achieve 13.69\% higher accuracy with the same memory usage, or equivalently, they are $13.25\times$ more memory-efficient at the same accuracy level compared to the baselines.

\subsection{Accuracy Heatmap : Dimensions and Memory Columns} 
Fig. \ref{fig:4} presents a heatmap of accuracy when varying the hyperparameters of our multi-centroid AM: dimensions and memory columns from 64 to 1024. This visualization allows us to optimize the AM structure according to the available hardware resources by adjusting dimensions and columns. Generally, higher dimensions correlate with improved accuracy, which is likely related to the quality of encoding. For MNIST and FMNIST, we observe that both higher dimensions and more columns (i.e., using more class vectors) correlate with improved accuracy. However, ISOLET dataset exhibits a different pattern, showing peak performance when using 128-256 columns. 

This divergence in optimal structure reflects the unique characteristics of each dataset. ISOLET, with its small sample size per class (approximately 240 samples per class) and large number of classes, shows peak performance with fewer columns. Using too many columns for ISOLET could lead to overfitting, potentially causing some class vectors to represent outliers rather than general class features. Conversely, MNIST and FMNIST (approximately 6000 samples per class), being larger and more diverse, benefit from additional class vectors without such overfitting concerns. These findings highlight the importance of adapting the AM structure to both the dataset characteristics and available computational resources.

\begin{figure}[t]
\centering
\includegraphics[width=1.0\linewidth]{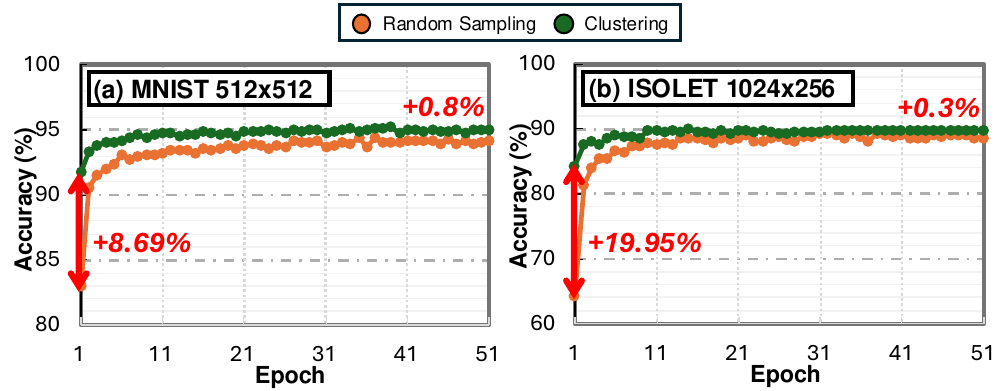}
\caption{Accuracy comparison between clustering and random sampling.}
\label{fig:5}
\vspace{-16pt}
\end{figure}

\begin{figure}[t]
\centering
\includegraphics[width=1.0\linewidth]{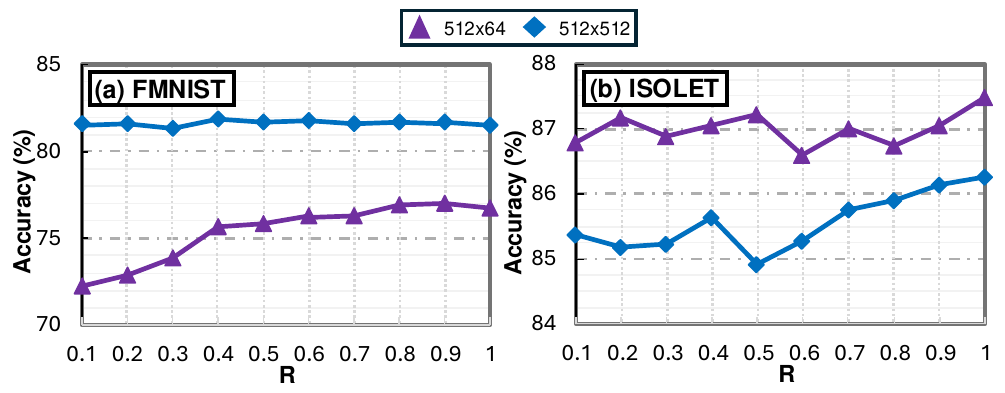}
\caption{Accuracy for different initial cluster ratios $R$ (0.1 to 1.0).}
\label{fig:6}
\vspace{-20pt}
\end{figure}

\begin{table*}[t] 
\centering
\renewcommand{\arraystretch}{1.03}
\setlength{\tabcolsep}{5pt}
\caption{Computation cycles, arrays and AM utilization for MNIST, FMNIST and ISOLET datasets using 128x128 IMC array.}
\begin{tabular}{|l|l|ccccc|ccccc|}
\hline
\multicolumn{2}{|c|}{\textbf{Dataset}} & \multicolumn{5}{c|}{\textbf{(a) MNIST, FMNIST}} & \multicolumn{5}{c|}{\textbf{(b) ISOLET}} \\
\hline
\multicolumn{2}{|c|}{\textbf{IMC Mapping}} & \multirow{2}{*}{\textbf{Basic}} & \multicolumn{2}{c}{\textbf{Partitioning}} & \multirow{2}{*}{\textbf{MEMHD}} & \multirow{2}{*}{\textbf{Improv.}} & \multirow{2}{*}{\textbf{Basic}} & \multicolumn{2}{c}{\textbf{Partitioning}} & \multirow{2}{*}{\textbf{MEMHD}} & \multirow{2}{*}{\textbf{Improv.}} \\
\cline{4-5} \cline{9-10}
\multicolumn{2}{|c|}{\textbf{Method}} & & P=5 & P=10 & & & & P=2 & P=4 & & \\
\hline
\multicolumn{2}{|c|}{\textbf{AM Structure}} & 10240x10 & 2048x50 & 1024x100 & \textbf{128x128} & - & 10240x26 & 5120x52 & 2560x104 & \textbf{512x128} & - \\
\hline
\multirow{3}{*}{\textbf{\# of Cycles}} & EM & 560 & 560 & 560 & \textbf{7} & \textbf{80$\times$ $\downarrow$} & 400 & 400 & 400 & \textbf{20} & \textbf{20$\times$ $\downarrow$} \\
& AM & 80 & 80 & 80 & \textbf{1} & \textbf{80$\times$ $\downarrow$} & 80 & 80 & 80 & \textbf{4} & \textbf{20$\times$ $\downarrow$} \\
& Total & 640 & 640 & 640 & \textbf{8} & \textbf{80$\times$ $\downarrow$} & 480 & 480 & 480 & \textbf{24} & \textbf{20$\times$ $\downarrow$} \\
\hline
\multirow{3}{*}{\textbf{\# of Arrays}} & EM & 560 & 560 & 560 & \textbf{7} & \textbf{80$\times$ $\downarrow$} & 400 & 400 & 400 & \textbf{20} & \textbf{20$\times$ $\downarrow$} \\
& AM & 80 & 16 & 8 & \textbf{1} & \textbf{8$\times$ $\downarrow$} & 80 & 40 & 20 & \textbf{4} & \textbf{5$\times$ $\downarrow$} \\
& Total & 640 & 576 & 568 & \textbf{8} & \textbf{71$\times$ $\downarrow$} & 480 & 440 & 420 & \textbf{24} & \textbf{17.5$\times$ $\downarrow$} \\
\hline
\textbf{Utilization} & AM & 7.81\% & 39.06\% & 78.13\% & \textbf{100\%} & \textbf{21.87\% $\uparrow$} & 20.31\% & 40.63\% & 81.25\% & \textbf{100\%} & \textbf{18.75\% $\uparrow$}\\
\hline
\end{tabular}
\vspace{-16pt}
\label{tab:2}
\end{table*}

\subsection{Clustering-based Initialization}
Our clustering-based initialization significantly outperforms random sampling in accuracy. Fig. \ref{fig:5} shows that for MNIST (512x512) and ISOLET (1024x256), clustering achieves 8.69\% and 19.95\% higher initial accuracies respectively, leading to improved final accuracies and faster convergence (within 10-20 epochs compared to 30-40 epochs for random sampling).

The initial clusters ratio $R$ influences the distribution of class vectors. 
As shown in Fig. \ref{fig:6}, $R$ has minimal effect at 512x512 but impacts performance at 512x64, with optimal values between 0.8 and 0.9. ISOLET achieves highest accuracy at $R=1.0$. These findings highlight the importance of proper initialization and $R$ selection in optimizing performance across datasets and $C$.

\subsection{Computation Cycle, Array Usage and AM Utilization}
In Fig. \ref{fig:4}, MEMHD demonstrated comparable or higher accuracy compared to BasicHDC with 10240D: our 128x128 achieved higher accuracy for both MNIST and FMNIST, while 512x128 showed comparable results for ISOLET. So, Table \ref{tab:2} compares our models with IMC baselines with 10240D when using 128x128 IMC array. We evaluate three key metrics: computation cycles, array usage, and AM utilization. Computation cycles refer to the number of operations performed when using a single array, while array usage indicates the number of arrays required to map the entire AM structure. AM utilization represents the ratio of actually mapped columns to the total columns in the IMC array. For partitioning, we consider two suitable scenarios for each dataset. Our model leverages lower dimensional vectors, which substantially reduces the EM's memory requirements. This reduction in dimensionality leads to fewer computation cycles and requires fewer IMC arrays, enhancing overall efficiency. Compared to the baseline, MEMHD achieves fully-utilization of AM at all times. As shown in Table \ref{tab:2}-(a), for MNIST and FMNIST datasets, our model demonstrates an 80× improvement in computational efficiency and requires 71× fewer arrays compared to the baseline. For ISOLET dataset, as presented in Table \ref{tab:2}-(b), our model achieves a 20× increase in computational efficiency and uses 17.5× fewer arrays.

\subsection{Energy Consumption and Cycles of AM}
\begin{figure}[t]
\centering
\includegraphics[width=0.9\linewidth]{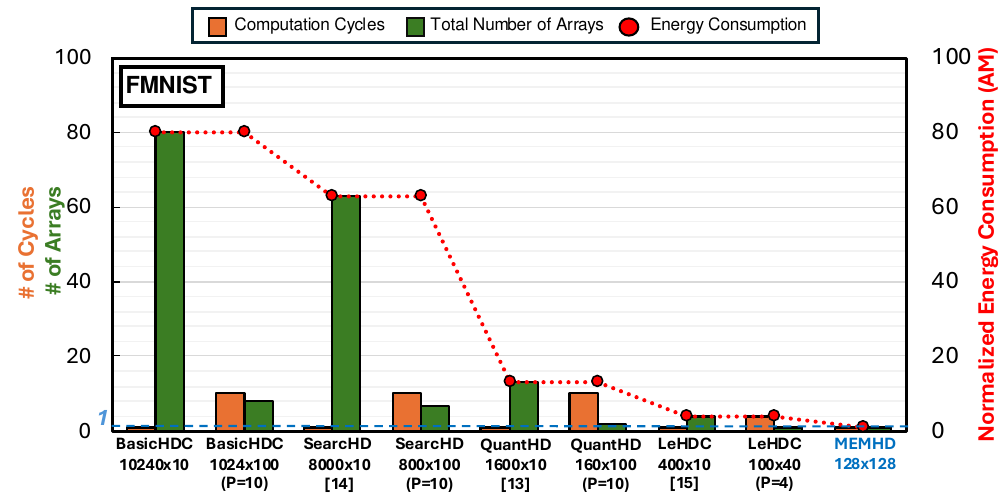}
\caption{Normalized energy of AM and cycles with array usage. MEMHD (128x128) achieves comparable accuracy to baselines with higher dimensions: BasicHDC (10240D), SearcHD (8000D), QuantHD (1600D), and LeHDC (400D).}
\label{fig:7}
\vspace{-20pt}
\end{figure}

Fig. \ref{fig:7} presents a comparative analysis of normalized AM energy consumption and cycles across all baseline models. While some baseline binary HDC models use ID-Level encoding, all models employ MVM-based associative search for inference, enabling a fair comparison of AM performance. The figure showcases models demonstrating equivalent accuracy to MEMHD 128x128 on the FMNIST dataset, as shown in Fig. \ref{fig:3}. The total number of arrays required to map the entire AM structure is proportional to the model's dimensionality. When the entire AM is mapped to arrays at once, models without partitioning can perform inference in a single cycle. However, for partitioned models, the number of cycles increases proportionally to the number of partitions. While partitioning strategies effectively reduce the number of required arrays, they proportionally increase the number of cycles, resulting in constant energy consumption across different partitioning schemes. MEMHD distinguishes itself by enabling associative search with just one computation in a single 128x128 array. This unique architecture allows MEMHD to achieve not only single-cycle inference but also significantly reduced energy consumption. Consequently, MEMHD demonstrates remarkable efficiency gains: it is $80\times$ more efficient than Basic HDC in terms of energy consumption. Moreover, compared to LeHDC, which represents the state-of-the-art in accuracy, MEMHD is $4\times$ more efficient.

\section{Conclusion} 
This paper introduced MEMHD, a Memory-Efficient Multi-centroid HDC framework that optimizes HDC for IMC architectures. MEMHD effectively trains multi-centroid AM through clustering-based initialization and quantization-aware iterative learning. As a result, this approach significantly outperforms binary HDC baselines, achieving up to 13.69\% higher accuracy with the same memory usage or $13.25\times$ more memory efficiency at the same accuracy level. By enabling full utilization of IMC arrays and one-shot (or few-shot) associative search, MEMHD reduces computation cycles by up to $80\times$ and array usage by up to $71\times$  compared to the partitioning method on 128×128 IMC arrays, substantially improving energy efficiency and cycles. These advancements pave the way for more efficient implementation of HDC in resource-constrained environments.

\section*{Acknowledgement}
This work was partly supported by the National Research Foundation of Korea (NRF) grant (No. RS-2024-00345732);
the Institute for Information \& communications Technology Planning \& Evaluation (IITP) grants (RS-2020-II201821, IITP-2021-0-02052, RS-2019-II190421, RS-2021-II212068); the Technology Innovation Program (RS-2023-00235718, 23040-15FC) funded by the Ministry of Trade, Industry \& Energy (MOTIE, Korea) grant (1415187505);
Samsung Electronics Co., Ltd (IO230404-05747-01); and the BK21-FOUR Project.

\vspace{12pt}
\color{red}
\end{document}